\documentclass[aps,notitlepage,showkeys,oneside,superscriptaddress,10pt]{revtex4-1}
\usepackage{amssymb}
\usepackage{float}

\begin{document}
\title{Supervisor of the Universe} 
\author{Victor A. Berezin}\thanks{berezin@inr.ac.ru}
\affiliation{Institute for Nuclear Research of the Russian Academy of Sciences, prospekt 60-letiya Oktyabrya 7a, Moscow 117312, Russia}
\author{Vyacheslav I. Dokuchaev}\thanks{dokuchaev@inr.ac.ru}
\affiliation{Institute for Nuclear Research of the Russian Academy of Sciences, prospekt 60-letiya Oktyabrya 7a, Moscow 117312, Russia}

\date{\today}

\begin{abstract}
In this paper, conformal invariant gravitation, based on Weyl geometry, is considered. In addition to the gravitational and matter action integrals, the interaction between the Weyl vector (entered in Weyl geometry) and the vector, representing the world line of the independent observer, are introduced. It is shown that the very existence of such an interaction selects the exponentially growing scale factor solutions among the cosmological vacuums.
\end{abstract}

\maketitle \tableofcontents

\section{Introduction}

In the present paper, following Roger Penrose \cite{Penrose,Penrose2,Penrose3} and Gerard\,‘t~Hooft \cite{Hooft,Hooft2,Hooft3}, it is suggested that the universe is conformal  invariant. Furthermore, it is suggested that the conformal invariace is described by Weyl geometry. The main problem of any conformal invariant gravitational theory is how to choose the “correct gauge”. G.\,’t Hooft \cite{Hooft3} proposed that different observers may see different pictures, i.e., different geometries. Evidently, this becomes possible only if the observer interacts somehow with the geometry. Furthermore, it is Weyl geometry that provides us with such a possibility. The observer is considered as being independent, i.\,e., the observer is not a dynamical variable and is not subject to variation. However, after the variation procedure, it is possible, in principle, to identify 
the observer with the matter flow. 

\section{Basics of Weyl Geometry and Weyl Conformal Gravity}

The elements of Weyl geometry \cite{Weyl,Weyl2} are the metric
tensor $g_{\mu\nu}$ (like in Riemann geometry) and the vector field $A_\mu(x)$, which is called the ``Weyl vector'' here on. 
The indices, denoted by Greek letters, take on the values 0 (time), 1, 2, 3 (space), $x$ are the observable four space-time
coordinates. The Weyl vector defines the covariant derivative of the metric tensor, namely,   
\begin{equation}
	\nabla_\lambda g_{\mu\nu}(x)=A_\lambda(x)g_{\mu\nu}(x).
	\label{Delta} 
\end{equation}
This relation leads to the connections provided
$\Gamma^\lambda_{\mu\nu}=\Gamma^\lambda_{\nu\mu}$:
\begin{equation}
	\Gamma ^\lambda_{\mu\nu}=C^\lambda_{\mu\nu}+W^\lambda_{\mu\nu},
	\label{Gamma} 
\end{equation} 
where $C^\lambda_{\mu\nu}$ are familiar Christoffel symbols,  
\begin{equation}
	C^\lambda_{\mu\nu}= \frac{1}{2}g^{\lambda\kappa}(g_{\kappa\mu,\nu}+g_{\kappa\nu,\mu}-g_{\mu\kappa,\nu})
	\label{Ch} 
\end{equation}  
and $W^\lambda_{\mu\nu}$ are 
\begin{equation}
	W^\lambda_{\mu\nu}=-\frac{1}{2}(A_\mu \delta^\lambda_\nu + A_\nu \delta^\lambda_\mu - A^\lambda g_{\mu\nu})\,.
	\label{WW} 
\end{equation}
Here, $\delta^\lambda_\nu$ is the Kronecker symbol (unit tensor), $g^{\lambda\kappa}$ is the inverse metric tensor defined by $g^{\lambda\kappa}g_{\kappa\nu}=\delta^\lambda_\nu$, and the comma "," denotes a coordinate partial derivative. The curvature tensor  is defined as: 
\begin{equation}
	R^{\mu}_{\phantom{\mu}\nu\lambda\sigma}=\frac{\partial \Gamma^\mu_{\nu\sigma}}{\partial x^\lambda}-\frac{\partial \Gamma^\mu_{\nu\lambda}}{\partial x^\sigma}+\Gamma^\mu_{\varkappa\lambda}\Gamma^\varkappa_{\nu\sigma}-\Gamma^\mu_{\varkappa\sigma}\Gamma^\varkappa_{\nu\lambda},
	\label{curvaturetensor} 
\end{equation} 
along with the convolutions, the Ricci tensor $R_{\mu\nu}=R^\lambda_{\phantom{\mu}\mu\lambda\sigma}$ and the curvature scalar $R^\lambda_\lambda$. Note that, in Weyl geometry, there are no familiar symmetry relations for the curvature and Ricci tensor.

In 1918, Hermann Weyl  constructed the unified theory of electromagnetic and gravitational fields \cite{Weyl}. He noticed that Maxwell equations are invariant under local conformal transformations (outside the sources) and claimed that the unified theory must posses  the same property. The local conformal transformation with the conformal factor $\Omega(x)$ is 
\begin{equation}
	ds^2=\Omega^2(x)d\hat s^2=  
	g_{\mu\nu}dx^\mu dx^\nu=\Omega^2(x)\hat g_{\mu\nu}dx^\mu dx^\nu;
	\label{conf} 
\end{equation}
which does not involve the coordinates, but the measures. If one demands $A_\mu$ to be the gauge field that transforms as: 
\begin{equation}
	A_\mu=\hat A_\mu+2\frac{\Omega_{,\mu}}{\Omega}\,, 	
	\label{confA} 
\end{equation}
then the connections  $\Gamma^\lambda_{\mu\nu}$ become conformal invariant,
\begin{equation}
	\Gamma^\lambda_{\mu\nu}=\hat\Gamma^\lambda_{\mu\nu}\,.	
	\label{confGamma} 
\end{equation}
Consequently, the curvature tensor and the Ricci tensor are also conformal invariant,
\begin{equation}
	R^\mu_{\phantom{1}\nu\lambda\sigma}=
	\hat R^\mu_{\phantom{1}\nu\lambda\sigma}, \quad R_{\mu\nu}=\hat R_{\mu\nu}\,.
	\label{confGamma2} 
\end{equation}
Evidently, one has yet another conformal invariant antisymmetric tensor,
\begin{equation}
	F_{\mu\nu}=A_{\nu,\mu} - A_{\mu,\nu}=\nabla_\mu A_\nu - \nabla_\nu A_\mu\,.
	\label{FW} 
\end{equation}
This was the reason why Weyl declared $A_\mu$ to be the electromagnetic vector potential, and $F_{\mu\nu}$ --- the electromagnetic strength tensor. In what follows, $A_\mu$ is considered as only a part of Weyl symmetry.

Following Weyl, let us write the gravitational action integral $S_{\rm  W}$ as
\begin{equation}
	S_{\rm  W}=\int\!{\cal L_{\rm W}}\sqrt{-g}\,d^4x\,,
	\label{ActionW} 
\end{equation}
\begin{equation}
	{\cal L_{\rm  W}}=\alpha_1  R_{\mu\nu\lambda\sigma}R^{\mu\nu\lambda\sigma}
	+\alpha_2R_{\mu\nu}R^{\mu\nu}+\alpha_3R^2+\alpha_4 F_{\mu\nu}F^{\mu\nu}\,,
	\label{LagrW} 
\end{equation}
which  is manifestly conformal invariant, since $\sqrt{-g}=\Omega^4\sqrt{-\hat g}$, $g$ is the determinant of the metric tensor. and $\alpha_1$, $\alpha_2$, $\alpha_3$, and $\alpha_4$ are constants.

\section{Setting the Problem}

When dealing with the conformal invariant theory, the most important question that readily arises is how to fix the gauge. The solution could be as follows \cite{Hooft,Hooft2,Hooft3,Penrose,Penrose2,Penrose3}. Different observers see the around world differently. To realize such a scenario, one needs an observer interacting somehow with the space--time geometry. Let us
choose a simplest possible interaction Lagrangian: 
\begin{equation}
	S_{\rm int}=\!\int\!\!{\cal L}_{\rm int}\sqrt{-g}\,d^4x,
	\label{intxi} 
\end{equation}
\begin{equation}
	{\cal L}_{\rm int}=-A_\mu \xi^\mu,
	\label{intxi2} 
\end{equation}
where $\xi^\mu$ describes the observer which is called, as in the title of this paper, "the supervisor''. An important note to be made: the supervisor $\xi^\mu$ is not a dynamical variable and, thus, is not subject to~variation. 

The total action integral,
\begin{equation}
	S_{\rm tot}=S_{\rm W}+S_{\rm int}+S_{\rm m},
	\label{Stot} 
\end{equation}
where $S_{\rm m}$ is the action integral for some matter fields. Another rather important note to be made: the gravitational part, $S_{\rm W}$, is conformal invariant, but the remaining parts, $S_{\rm int}$ and $S_{\rm m}$, do not need to have the same property. However, the variation, $\delta S_{\rm int}+\delta S_{\rm m}$, {\it does.}

The conformal invariance means that, under the variations of conformal factor, $\Omega$, 
\begin{equation}
	\frac{\delta S_{\rm tot}}{\delta\Omega}=0.
	\label{deltaStot} 
\end{equation}
Here, $\delta\Omega$ enters both $\delta \sqrt{-g}=2\Omega\hat g_{\mu\nu}\delta\Omega=2(\delta\Omega/\Omega)g_{\mu\nu}$ and $\delta A_\mu=2\delta(\Omega_{,\mu}/\Omega)$. Hence,
\begin{equation}
	\delta S_{\rm int}=-2\!\int\!\xi^\mu\left(\frac{\delta \Omega}{\Omega}\right) \sqrt{-g}\,d^4x -\!\int\!(A_\mu \xi^\mu)g^{\nu\lambda}g_{\nu\lambda} \left(\frac{\delta \Omega}{\Omega}\right)\sqrt{-g}\,d^4x,
	\label{intxi3} 
\end{equation}
then
\begin{equation}
	\delta S_{\rm int}=2\xi^\mu_{;\mu}-4(A_\mu \xi^\mu).
	\label{intxi4} 
\end{equation}

Further on, by definition %, 
($G^\mu=-\delta S_{\rm int}/\delta A_\mu$):
\begin{equation}
	\delta S_{\rm m}=-\frac{1}{2}\int\!T^{\mu\nu}(\delta g_{\mu\nu}) \sqrt{-g}\,d^4x -\!\int\!G^\mu(\delta A_\mu)\sqrt{-g}\,d^4x + \!\int\!\frac{\delta{\cal L_{\rm W}}}{\delta\Psi}(\delta\Psi) \sqrt{-g}\,d^4x,
	\label{delta5} 
\end{equation}	
where $\Psi$ is the collective matter fields variable, and $T^{\mu\nu}$ is the energy-momentum tensor. Hence,
\begin{equation}
	\frac{\delta S_{\rm m}}{\delta\Omega}=2G^\mu_{;\mu}-{\rm Trace}(T^{\mu\nu}).
	\label{delta4} 
\end{equation}	
Finally, one obtains the following self-consistency condition:
\begin{equation}
	2(\xi^\mu_{;\mu}+G^\mu_{;\mu})=4A_\mu\xi^\mu+T^\mu_\mu\, ,
	\label{intxi2b} 
\end{equation}
from where the "Weyl current'' $G^\mu$  may come. To clarify this, let us consider the specific example, namely, the perfect fluid. The perfect fluid consists of particles interacting with each other directly and gravitationally. In Riemann geometry, the gravitational interaction is due to the metric tensor field only. Meantime, Weyl geometry involves, in addition, the Weyl vector $A_\mu$. Therefore, the additional invariant, $B=A_\mu u^\mu$, exists for describing the gravitational interaction of the particle with the four-velocity $u^\mu$ \cite{bde16,bdes21}.

The action integral for the perfect fluid in Riemann geometry can be written as \cite{Ray,Berezin}:
\begin{eqnarray}
	S_{\rm m} &=&  -\!\!\int\!\varepsilon(X,n)\sqrt{-g}\,d^4x + \!\!\int\!\lambda_0(u_\mu u^\mu-1)\sqrt{-g}\,d^4x \nonumber \\ 
	&&+\!\!\int\!\lambda_1(n u^\mu)_{;\mu}\sqrt{-g}\,d^4x + \!\!\int\!\lambda_2 X_{,\mu}u^\mu\sqrt{-g}\,d^4x \, ,
	\label{SmPerfect} 
\end{eqnarray}
with the dynamical variables: $n$, $u^\mu$, and $X$. In this action integral $n(x)$ being the invariant particle number density, $X$ being the auxiliary variable numbering the particle trajectories, $\varepsilon$ is the invariant energy density,  
$\lambda_0$, $\lambda_1$, and $\lambda_2$ are  the Lagrange multipliers, and the semicolon ";'' denotes the covariant derivative with metric connections (Christoffel symbols).

The Lagrange multipliers, $\lambda_i$, provide us with the constraints: the normalization condition for four velocities, particle number conservation  and the constant value  of $X$ along the trajectories. Evidently, one can easily insert our new invariant $B$ into $\varepsilon$:  $\varepsilon(X,n)\; \rightarrow \; \varepsilon(X,B,n)$. It can be shown that the constraint $(n u^\mu)_{;\mu}=0$ is not conformal invariant. Indeed, by construction (see,  e.g., \cite{LL}), $g_{\mu\nu}$ is transformed under the conformal transformation,
$g_{\mu\nu}=\Omega^2(x)\hat g_{\mu\nu}$ and, respectively,
\begin{equation}
	n=\frac{\hat n}{\Omega^4}, \quad u^\mu=\frac{1}{\Omega^4}\hat u^\mu, \quad \sqrt{-g}=\Omega^4\sqrt{-\hat g}.
	\label{intxi4b} 
\end{equation}
Therefore,
\begin{eqnarray}
	(n u^\mu)_{;\mu}\sqrt{-g}&=&(n u^\mu\sqrt{-g})_{,\mu}=\left(\frac{\hat n}{\Omega}  \hat u^\mu\sqrt{-\hat g}\right)_{,\mu} \nonumber \\ 
	&=&\frac{1}{\Omega}(\hat n\sqrt{-\hat g}u^\mu)_{,\mu} -\frac{\Omega_{,\mu}}{\Omega^2}(n\sqrt{-\hat g}\hat u^\mu).
	\label{nu4} 
\end{eqnarray}
At the same time, if $(n u^\mu\sqrt{-g})_{,\mu}=0$, then $(\hat n \hat  u^\mu\sqrt{-\hat  g})_{,\mu}=0$ as well, because the number of particles can simply be counted. Thus, one can see that {\it the conformal invariance requires the particle number conservation.}

Let us choose the simplest possible form for the particle creation rate: $(n u^\mu)_{;\mu}=\Phi(B,n)$. Then, the action integral for the perfect fluid in the Weyl geometry reads:
\begin{eqnarray}
	S_{\rm m}&=&-\!\!\int\!\!\varepsilon(X,\varphi(B)n)\sqrt{-g}\,d^4x + \!\!\int\!\lambda_0(u_\mu u^\mu-1)\sqrt{-g}\,d^4x \nonumber \\ 
	&+&\int\lambda_1\left((\varphi_1(B)n u^\mu)_{;\mu}-\Phi(B,n)\right)\!\sqrt{-g}\,d^4x +\int\!\lambda_2 X_{,\mu}u^\mu\sqrt{-g}\,d^4x.
	\label{Sm3} 
\end{eqnarray}

Not all the field equations and equations of motion in general form are given here, being postponed, to Sect.~\ref{Sect4} of cosmological applications. Here, only the expressions for $G^\mu$ and $T^{\mu\nu}$ are given:
\begin{equation}
	G^\mu=\frac{\partial\varepsilon}{\partial B}u^\mu +\lambda_1\frac{\partial\Phi}{\partial B}u^\mu,
	\label{Gmu} 
\end{equation}
\begin{eqnarray}
	T^{\mu\nu}&=&\left\{ (\varepsilon+p)+ \lambda_1(\Phi+\Pi) -B\left(\frac{\partial\varepsilon}{\partial B} +\lambda_1\frac{\partial\Phi}{\partial B}\right) \right\} u^\mu u^\nu \nonumber
	\\
	&&+\left\{p+\lambda_1\Pi\right\} g^{\mu\nu},
	\label{Tmunu5} 
\end{eqnarray}
where the hydrodynamical pressure, $p=n\partial\varepsilon/\partial n -\varepsilon$, and %the 
$\Pi=n\partial \Phi/\partial n -\Phi$.

\section{Application to Cosmology}
\label{Sect4}

By cosmology, we understand any homogeneous and isotropic space--time manifold with the Robertson--Walker metric:
\begin{equation}
	ds^2=dt^2-a^2(t)\left(\frac{dr^2}{1-kr^2}+r^2(d\theta^2+\sin^2\theta d\varphi^2) \right),
	\label{RW} 
\end{equation}
where $k=0,\pm1$, $a(t)$ is the scale factor, and $t$ is the cosmological time. Because of the symmetry, the Weyl vector $A_\mu$ 
has one nonzero component only, depending on the cosmological time, $A_\mu=(A_0,0,0,0)$. It follows then, that
\begin{equation}
	F_{\mu\nu}\equiv0.
	\label{basic1} 
\end{equation}
Moreover, by choosing a suitable conformal factor, one can always set  $A_\mu=0$, and 
\begin{equation}
	B=A_\mu u^\mu=0.
	\label{basic2} 
\end{equation}
The solutions in this special gauge will be called "the basic solutions''. 

Let us introduce the following notations:
\begin{eqnarray}
	\varepsilon_0(X,n)&=&\varepsilon (X,0,n),
	\\
	%\frac{\partial\varepsilon}{\partial B}(X,0,n)&=& \varepsilon_1(X,n), \\ \Phi(0,n)&=&\Phi_0(n), \\
	\varepsilon_1(X,n)&=&\frac{\partial\varepsilon}{\partial B}(X,0,n), \\ \Phi_0(n)&=&\Phi(0,n), \\
	%\frac{\partial\Phi}{\partial B}(0,n)&=& \Phi_1(n).
	\Phi_1(n)&=& \frac{\partial\Phi}{\partial B}(0,n).
	\label{Tmunu4} 
\end{eqnarray}
The equations of motion for the perfect fluid can be reduced to the only equation:
\begin{equation}
	-(\varepsilon_0+p_0)-\lambda_1(\Phi_0+\Pi_0)-n\dot\lambda_1=0,
	\label{final1} 
\end{equation}
($u^0=1$, $u^i=0$, $X={\rm const}$), and the remaining constraint is:
\begin{equation}
	\frac{(n a^3)^{\dot{}}}{a^3}=\Phi_0(n).
	\label{lambda1} 
\end{equation}
The energy--momentum tensor, $T^\nu_\mu=(T^0_0,T^1_1=T^2_2=T^3_3 )$, is:
\begin{eqnarray}
	T^0_0&=&\varepsilon_0+\lambda_1\Phi_0, 	\label{Tconserv2}  \\
	T^1_1&=&-(p_0+\lambda_1\Pi_0). 	\label{Tconserv3} 
\end{eqnarray}
The Weyl current, $G^\mu$, has only one nonzero component,
\begin{equation}
	G^0=\varepsilon_1+\lambda_1\Phi_1,
	\label{Gmu2} 
\end{equation}
and the self-consistency condition becomes: 
\begin{equation}
	2\left(\frac{(\xi^0a^3)^{\dot{}}}{a^3}+\frac{(G^0a^3)^{\dot{}}}{a^3}\right)=
	\varepsilon_0+\lambda_1\Phi_0-3(p_0+\lambda_1\Pi_0).
	\label{G0a3} 
\end{equation}

The gravitational field equations  are divided into the vector and tensor parts. The vector equations are reduced to
\begin{equation}
	-6\gamma \dot R=\xi^0+G^0, 	\label{final3}
\end{equation}
while tensor equations become: 
\begin{eqnarray}
	-12\gamma\left\{\frac{\dot a}{a}\dot R+ R\left(\frac{R}{12}+\frac{\dot a^2+k}{a^2}\right)\right\}&=&T^0_0, 
	\label{final4} \\
	-4\gamma\left\{\ddot R+2\frac{\dot a}{a}\dot R
	- R\left(\frac{R}{12}+\frac{\dot a^2+k}{a^2}\right)\right\} &=& T^1_1.
	\label{final5}
\end{eqnarray}
Here, $\gamma=(\alpha_1+\alpha_2+3\alpha_3)/3$, the dot defines the time-derivative, and $R$ is the curvature scalar:  
\begin{equation}
	R=-6\left(\frac{\ddot a}{a}+\frac{\dot a^2\!+\!k}{a^2}\right). 
	\label{final2} 
\end{equation}

One can show that the self-consistency condition is, actually, the  consequence of the field equations  (\ref{final3})--(\ref{final5}). Note also that in the special gauge used here, the supervisor does not enter the right hand side of the tensor equations. In total, there are five equations for seven unknown functions.

\section{Basic Solutions}

Let us start with vacuum solutions. The main question one would like to answer is whether or not the supervisor  may exist in the absence of matter fields, or in other words, as "the ethereal ghost''.

Since in the vacuum $T^0_0=T^1_1=\varepsilon_0=\Phi_0=\Phi_1=G^0=0$, one gets
$\xi^0=-6\gamma b/a^3$, $b={\rm const}$, and
\begin{eqnarray}\label{T00b}
	&&\dot R=\frac{b}{a^3}, \\
	&&\frac{\dot a}{a}\dot R+ R\left(\frac{R}{12}+\frac{\dot a^2+k}{a^2}\right)=0, \\
	&&R=-6\left(\frac{\ddot a}{a}+\frac{\dot a^2+k}{a^2}\right)=0.
\end{eqnarray}
Fortunately, this seemingly overdetermined set of equations has the following~solution:
\begin{equation}\label{solution}
	az\frac{dz}{da}=\pm\sqrt{(z^2+k^2)-\frac{1}{3}zb},
\end{equation}
where $x(a)=\dot a$. No detailed investigation of  Equation~\ref{solution}  to be given here. Two points are of most importance here. 

\begin{enumerate}
	\item The supervisor survives in the vacuum.
	\item For large enough $a$ \ $(a \gg a_0)$, 
	\begin{equation}\label{largea}
		\dot a=\left(\frac{|b|}{12}\right)^{1/3}\frac{a}{a_0},
	\end{equation} 
\end{enumerate}
i.\,e., one obtains the exponential growth (due to the {\it presence of the supervisor}). In the same regime, the curvature scalar tends to the limiting value, 
\begin{equation}\label{limiting}
	R_0=-12^{1/3}\frac{|b|^{2/3}}{a_0^{\phantom{a}2}}<0.
\end{equation}

One can show that the basic solutions with a nonzero but traceless energy--momentum tensor will have the same properties, i.\,e., the solutions exponentially expand for large enough scale factors. Furthermore, the same is true in quite general cases 
provided the invariant particle number density $n(a)$ is an decreasing function.

\section{Conclusions}

In the present paper, an attempt to introduce an outside (external) observer in the least action integral, called "the supervisor', is made. Since such an observer is naturally described by some world line, the simplest geometry for 
the observer incorporation appeared to be Weyl geometry, which contains both the metric tensor and the vector field. 

The main feature of the presented approach  is that the supervisor is not the dynamical variable and, then, is not subject to the variation. After the variation procedure, it can be chosen freely and even identified with all the inhabitants of the universe. Here, only the cosmological applications is considered.

It appeared that on cosmological scale, one can always find an appropriate conformal factor that makes the Weyl gauge field  
equal zero. The corresponding solutions called here the basic ones and it is found that the scale factor for all of the solutions 
exhibit the exponential growth. The same is true for the non-vacuum solutions with a traceless energy--momentum tensor as well as  in the more general case of the perfect fluid, when the invariant particle number density is a decreasing function of the scale factor. 

No data are pretended to be explained here. The aim of this study is just to study some cosmological features of Weyl geometry by presenting the exact vacuum solution. In regard to the reasons and motivations for introducing Weyl geometry and a new concept, “the supervisor of the Universe”, let us to clarify the aim of this paper.

Only Weyl geometry is considered which differs radically from the Enstein conformal gravity, and it is demonstrated that Weyl geometry permits the existence of the independent observer (“the supervisor”) at least in the very early Universe. The word “independent” is used here in order to emphasize that the vector describing the “observer”  is not subject to variation. It can be identified physically with some matter flow, but already in the field equations.

Any further detailed investigation is postponed for the future, especially the for case when Maxim Khlopov, whose 70th anniversary we are celebrating now, may take part as one of the supervisors.

%\acknowledgements{We are grateful to E.\,O.\,Babichev, Yu.\,N.\,Eroshenko and A.\,L.\,Smirnov for stimulating discussions.}

\acknowledgments{We are grateful to E.\,O.\,Babichev, Yu.\,N.\,Eroshenko and A.\,L.\,Smirnov for stimulating discussions.}

\end{document}